\documentstyle[paspconf,epsf]{article}

\begin{document}

\title{Galaxy Morphology in
the GTO-NICMOS Northern Hubble Deep Field}

\author{Andrew J.~Bunker\altaffilmark{1}}
\affil{NICMOS Postdoctoral Research Fellow, Dept.\ of Astronomy,
U.C.\ Berkeley, 601 Campbell Hall, Berkeley CA94720\ \ {\tt
email:~bunker@bigz.berkeley.edu}} 
\altaffiltext{1}{In collaboration with Hyron Spinrad (Department of
Astronomy, U.C.\ Berkeley),
Rodger Thompson (Steward Observatory, University of Arizona) and the
NICMOS-GTO team.}
 
\begin{abstract}
The increased incidence of morphologically peculiar galaxies at faint
magnitudes in the optical could be attributable either to
``morphological $k$-corrections'' (the change in appearance when viewing
high-$z$ objects at shorter rest-frame wavelengths), or an increase in
the incidence of truly irregular systems with redshift.  The deep,
high-resolution GTO-NICMOS near-IR imaging of a portion of the northern
Hubble Deep Field has been combined with the WFPC\,2 data and
photometric redshift estimates to study the redshift evolution of
morphology, comparing galaxy appearance at the same rest-wavelengths
(Bunker, Spinrad \& Thompson 1999). It appears that morphological
$k$-corrections are only significant in a minority of cases, and that
once these are accounted for, evolution is still demanded -- galaxies
were smaller and more irregular in the past, with some of the
peculiarities probably merger-related. This multi-waveband data set also
enables a study of the spatially-resolved stellar populations in distant
galaxies.  A near-infrared analysis of some of the brighter spirals
shows more pronounced barred structure than in the optical, indicating
that the apparent decline in barred spirals at faint magnitudes in the
optical HDF may be due to band-shifting effects at the higher redshifts,
rather than intrinsic evolution.
\end{abstract}


\section{Introduction}

When coupled with distance estimates such as photometric redshifts,
the study of morphology has the potential to probe the dynamical state
and evolution of galaxies. However, morphological classification is only
reliable to redshifts of a few tenths when hampered by ground-based
seeing.  The deep, high-resolution WFPC\,2 imaging of the Hubble Deep
Field (HDF, Williams et al.\ 1996) dramatically pushed the study of
galaxy morphology to faint magnitudes and high redshifts, revealing that
by $I_{AB}>24$, the traditional Hubble sequence no longer provides an
adequate description of most galaxies (Abraham et al.\ 1996).

\begin{figure}[ht]
\plottwo{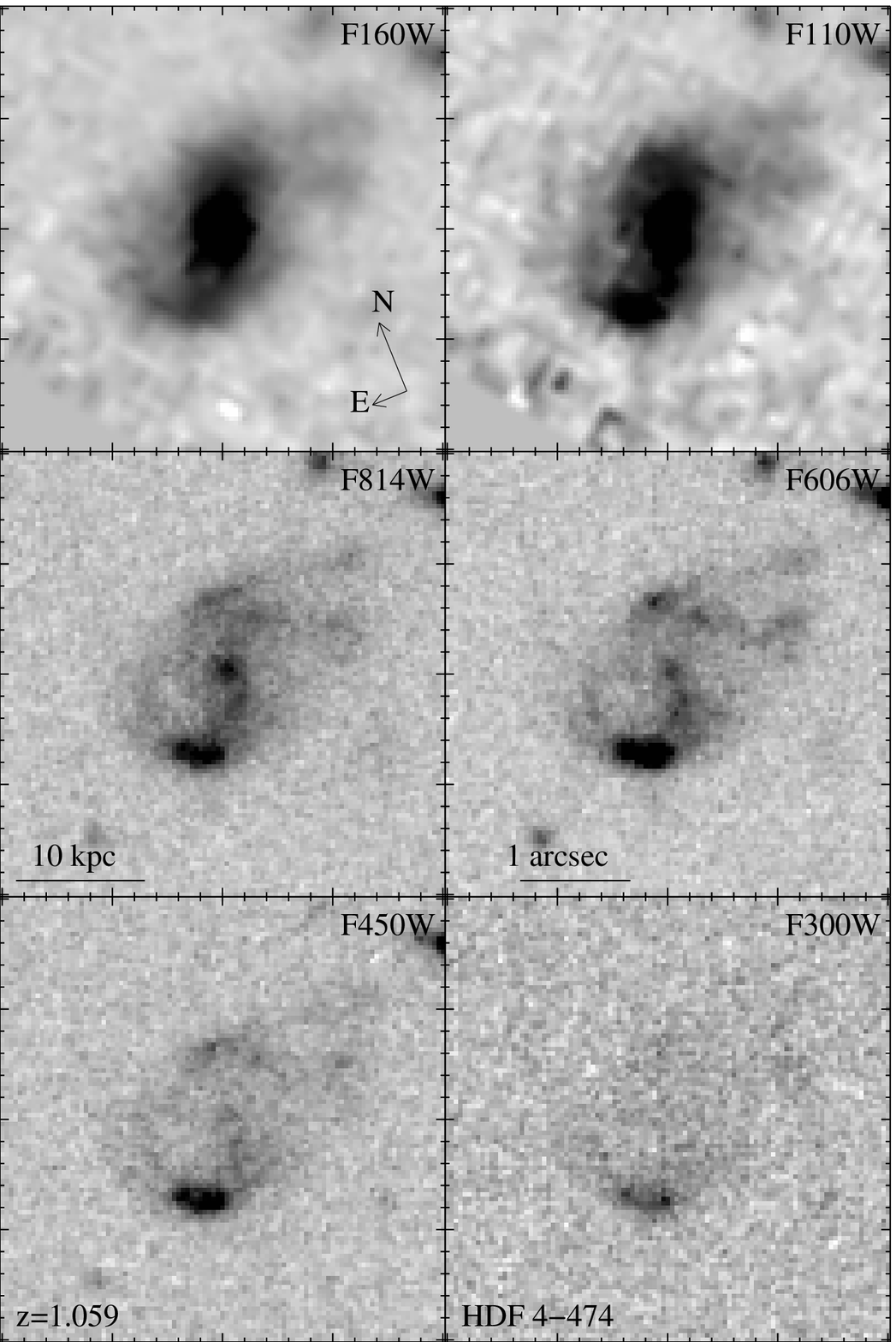}{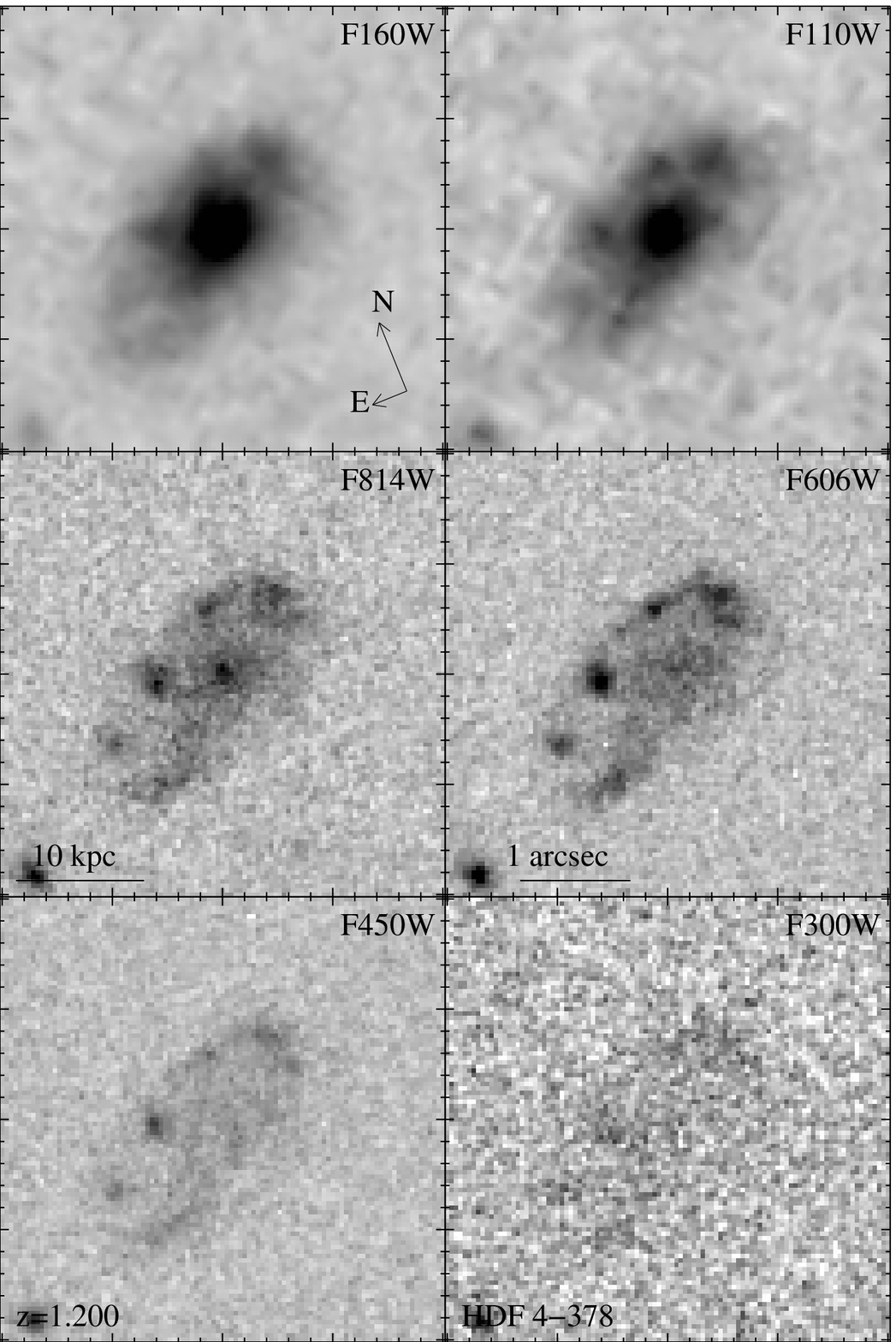}
\caption{Spiral galaxies at $z\approx 1$, showing the great change in
apparent morphology going from the optical (the rest-UV, where the
appearance is irregular) to the near-IR, where their true spiral nature
is revealed. In the case of HDF\,4-474 (left), the WFPC\,2 images are
dominated by a star forming knot, and for HDF\,4-378 (right) the
older/redder population of the bulge is only visible at IR wavelengths.}
\label{fig:spirals}
\plottwo{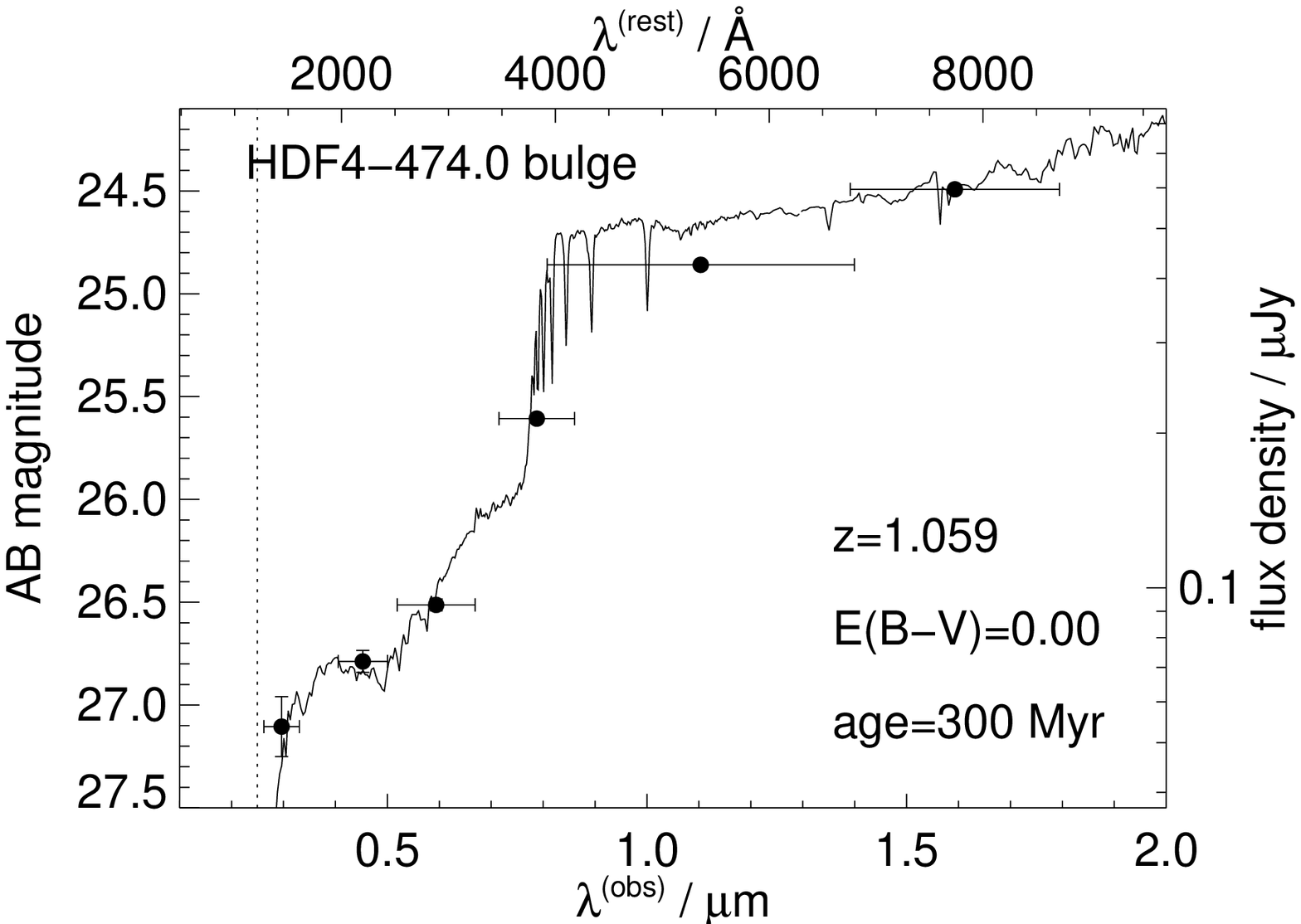}{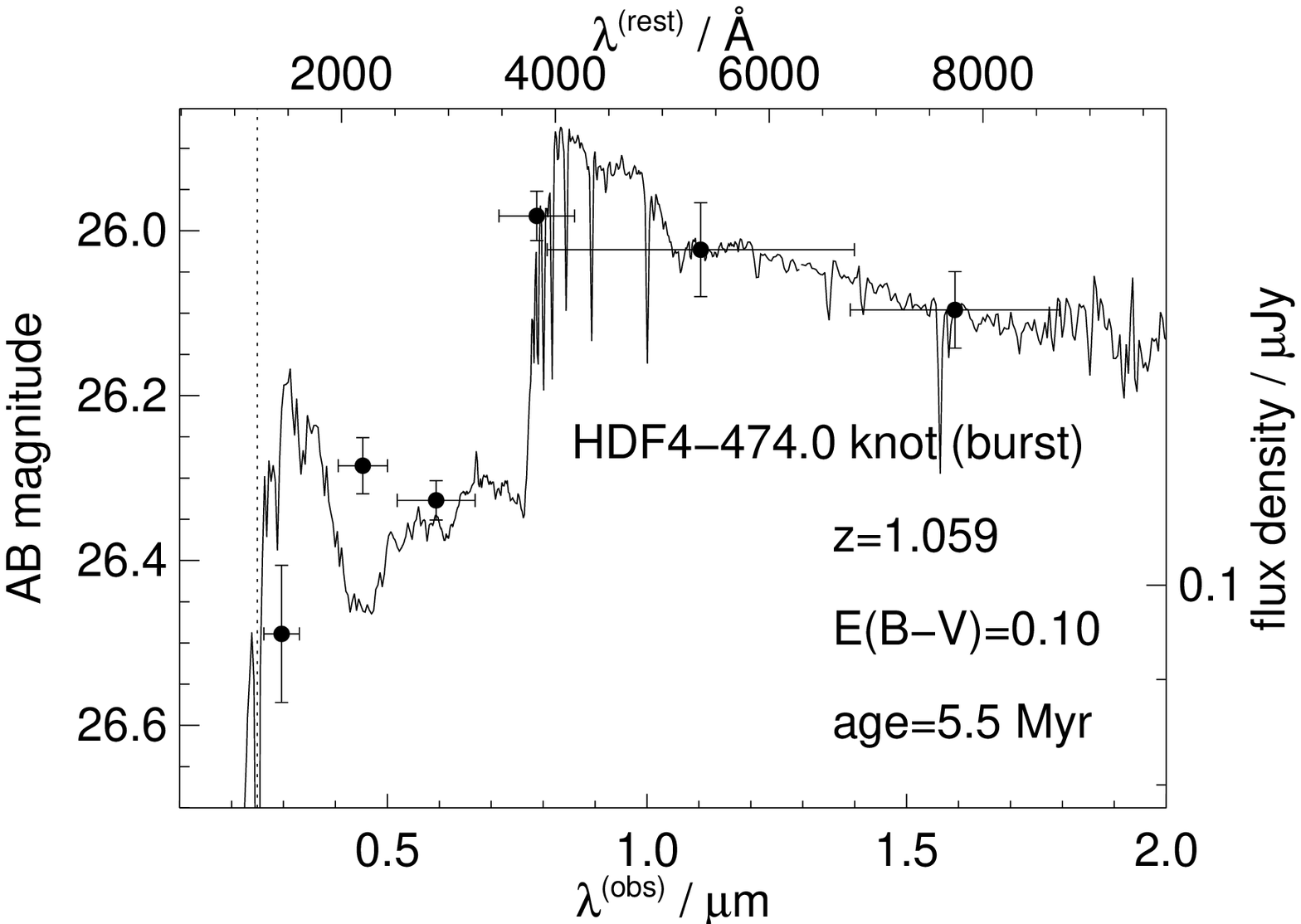}
\caption{Stellar population fits to two spatially resolved areas of the
$z\approx 1$ spiral HDF\,4-474 (see Fig.~\ref{fig:spirals}), using the
latest version of the Bruzual \& Charlot (1993) models. The bulge (left
panel) is clearly very much older than the star-forming H{\scriptsize
II} region in one of the spiral arms (right panel).}
\label{fig:spiralSEDs}
\end{figure}

Some of the faint peculiar galaxies are sub-luminous irregulars at
modest redshifts, while others are higher-$z$. But are they ``true
peculiars'' -- the counterparts to local irregulars?  Matters are
complicated by {\em morphological $k$-corrections:} for single
waveband selection, shorter rest wavelengths are sampled in higher-$z$
galaxies. The rest-UV is dominated by sites of recent star formation,
and it is known that the appearance of local Hubble-sequence galaxies
can be very different in the UV compared to the optical (e.g., O'Connell
1997). This change in apparent morphology, resulting from a dispersion of
stellar populations, is well illustrated by some HDF spirals at moderate
redshift ($z\approx 1$) which undergo a complete metamorphosis from the
observed optical to the near-IR (Fig.~\ref{fig:spirals}).

To address whether the apparent increased incidence of peculiars in the
optical at faint magnitudes is attributable to genuine evolution in
the fraction of irregular galaxies, or whether it is predominantly due
to band-shifting effects, the appearance of galaxies over a variety of
redshifts should be compared at the same rest wavelength
(Fig.~\ref{fig:montage}).

\section{A Study of HDF-North}

\begin{figure}[ht]
\plotone{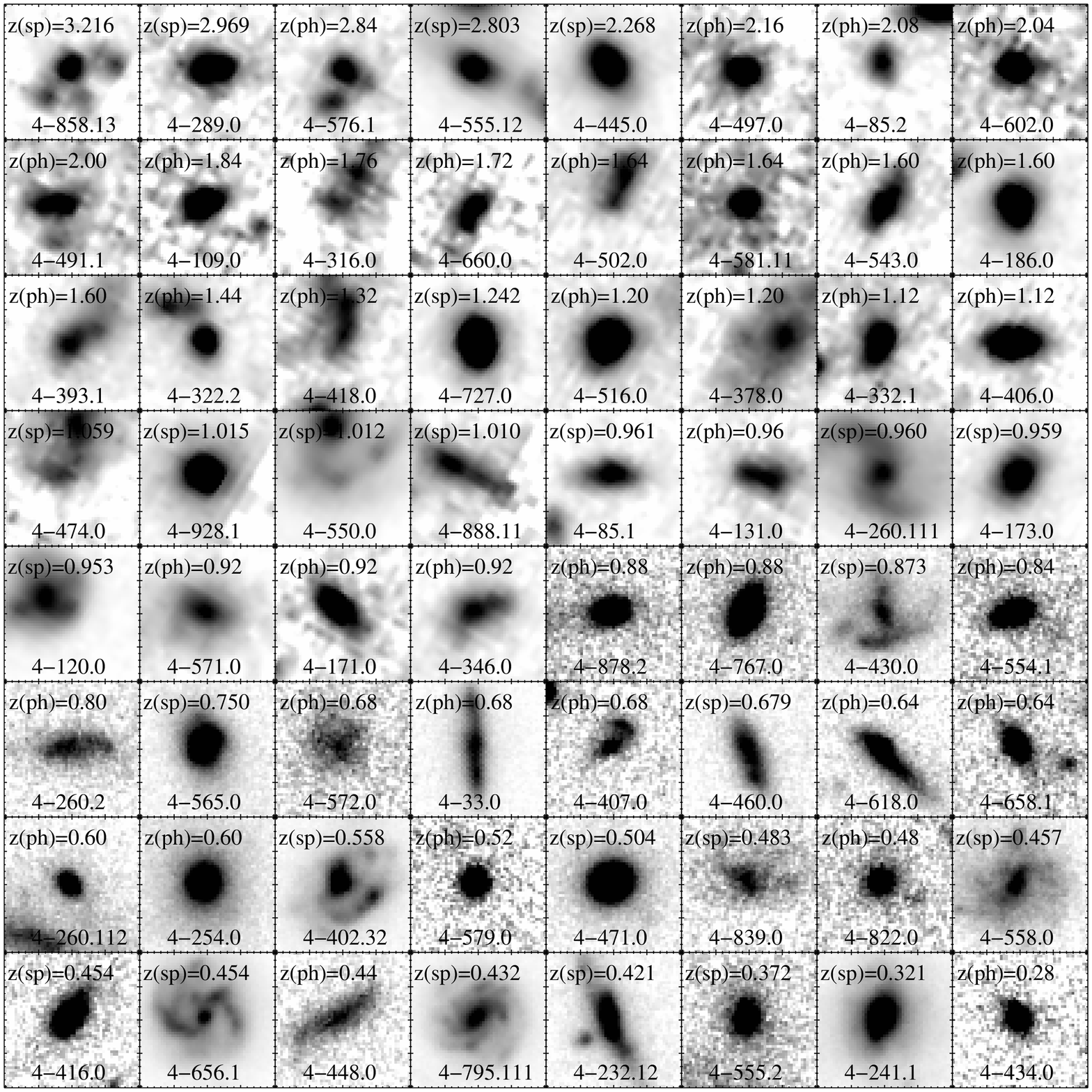}
\caption{A montage of galaxies down to $I_{AB}<26.0$ which lie within
the GTO-NICMOS Hubble Deep Field, ranked in order of redshift. Each
galaxy is displayed in the waveband which most closely matches the
rest-frame $B$-band, from the WFPC\,2 images (F450W--'$B$',
F606W--'$V$', F814W--'$I$') and the NIC\,3 data (F110W--'$J$',
F160W--'$H$'). Each box is 2\,arcsec across. At higher redshifts, the
incidence of the familiar Hubble-sequence galaxies declines greatly. The
identification numbers come from the catalog of Williams et al.\ (1996),
and the photometric redshifts are taken from Fern\'{a}ndez-Soto,
Lanzetta \& Yahil (1999). Those denoted by `z(sp)' have
spectroscopically-determined redshifts.}
\label{fig:montage}
\end{figure}

We have analyzed galaxy morphology to faint magnitudes in HDF-North
using the optical \& near-IR HST images (Bunker, Spinrad \& Thompson
1999). We have studied the GTO-NICMOS data set (Thompson et al.\ 1999),
a 1\,arcmin$^{2}$ area of the HDF imaged for 49 orbits with NIC\,3 in
both F110W ($\approx J$-band) and F160W ($\approx H$-band). Combined
with the four WFPC\,2 pass-bands (Fig.~\ref{fig:montage}), this data set
provides deep, multi-color, high-resolution imaging extending out to
1.6\,$\mu$m -- the rest-optical at $z\sim 2$.  We use the redshifts of
the galaxies to match the rest-wavelengths, determine intrinsic
luminosities, and to fit stellar populations/dust reddening to the
spectral energy distributions. Where available, we use the
spectroscopically-measured redshifts (from Cohen et al.\ 1996 unless
otherwise noted). Where no published spectroscopic redshift exists, we
adopt the photometric redshift estimate of Fern\'{a}ndez-Soto, Lanzetta
\& Yahil (1999).

\subsubsection{Comparative Morphology:}

Down to $I_{AB}\approx 26$ (the brightest 100 galaxies in GTO-NICMOS
field):
only about 1/6 of galaxies change their appearance
greatly between the WFPC\,2 and NICMOS images -- these have large
morphological $k$-corrections;
about half of the galaxies retain the same
morphology in all 
wavebands (above the redshifted Lyman break) and are ``true irregulars'';
the remaining third of galaxies are too compact
for changes in 
morphology to be ascertained (the NIC\,3 PSF has a FWHM of $\approx
0.25$arcsec); for most cosmologies, the higher-redshift systems are on
average more compact.

\subsubsection{Spatially-Resolved Stellar Populations:}

\begin{figure}[ht]
\plottwo{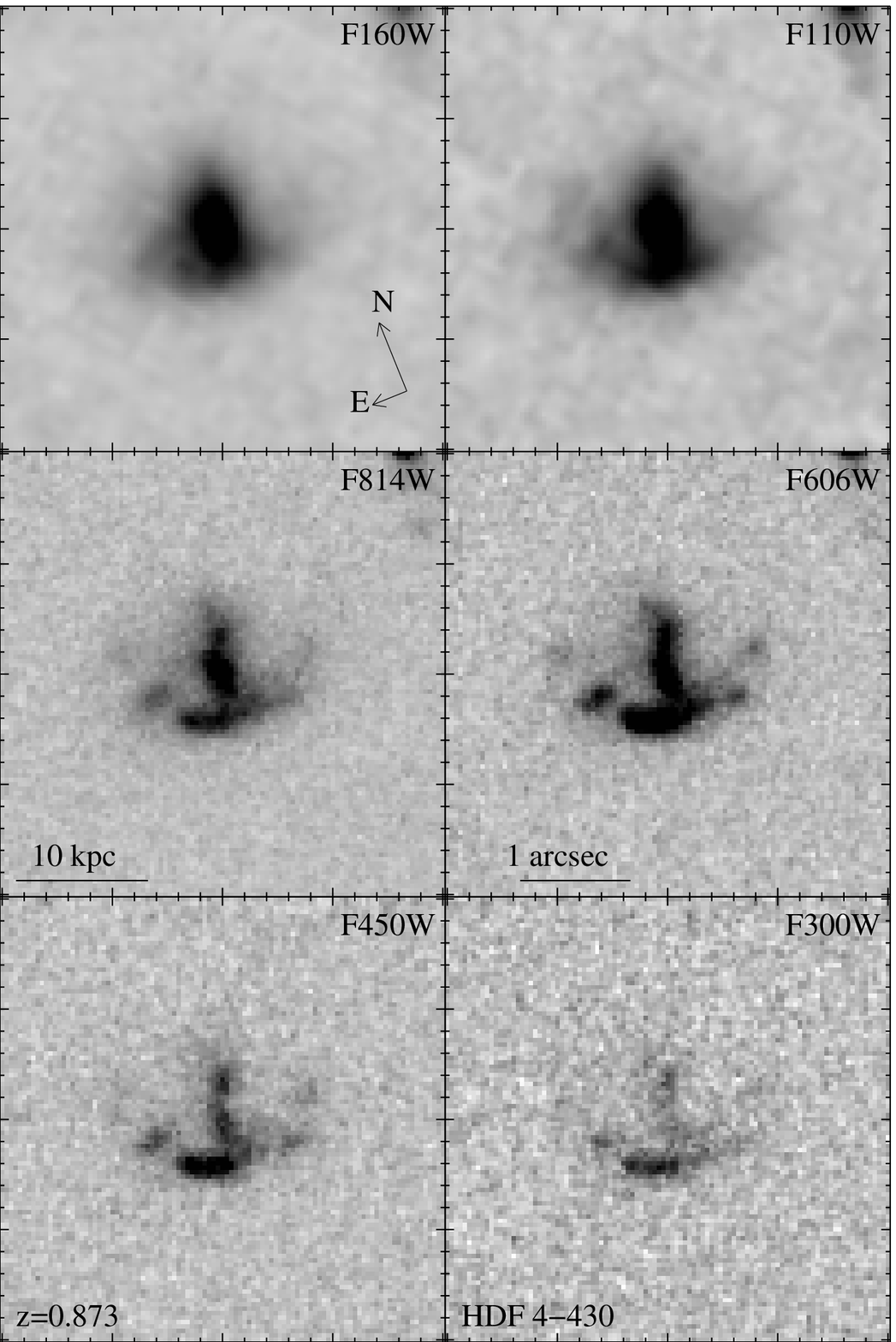}{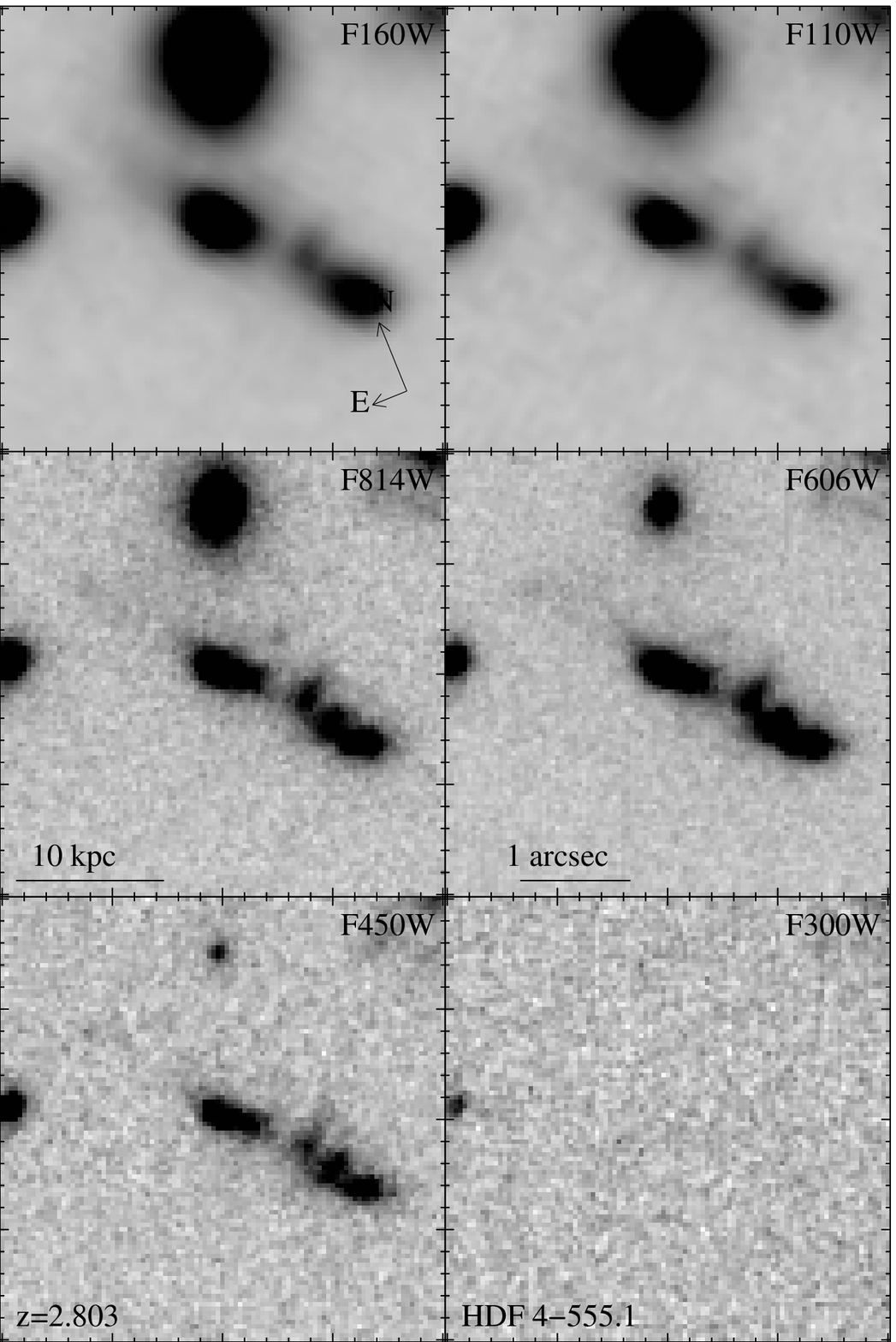}
\caption{Examples of a
bow-shock interacting system (left) and a chain galaxy (right). Note the
bow-shock area itself is comparatively blue, implying a young stellar
population with star formation presumably triggered by the shock front,
whereas the redder (older) core of the galaxy is more prominent in the
near-IR. The chain galaxy (the two-component $U$-drop  called ``the
Hot Dog''; Steidel et al.\ 1996, Bunker et al.\ 1998) appears the
same at all wavelengths and is blue, implying a relatively homogeneous,
young population (a prim\ae val galaxy candidate?).}
\label{fig:im430and555}
\plottwo{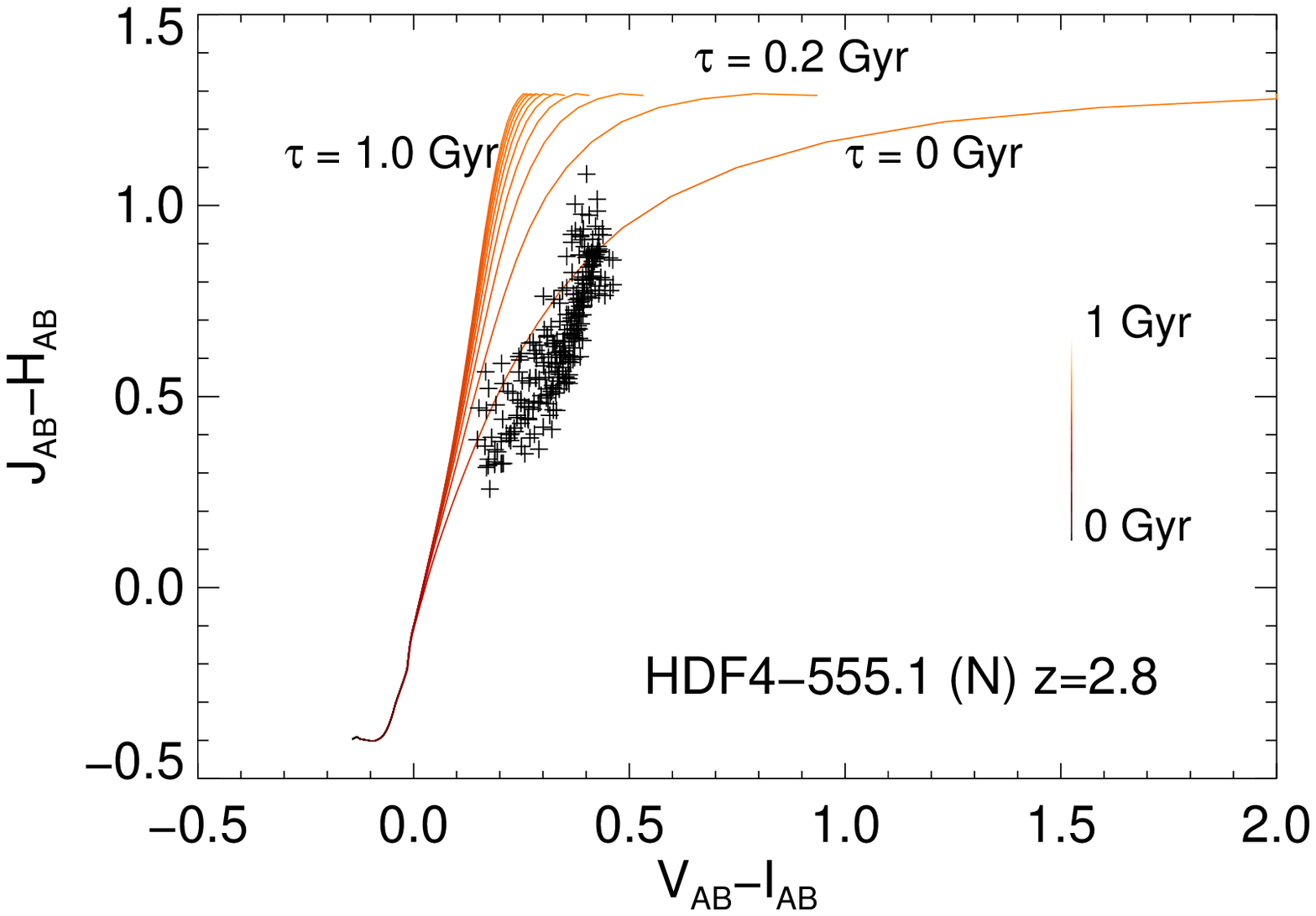}{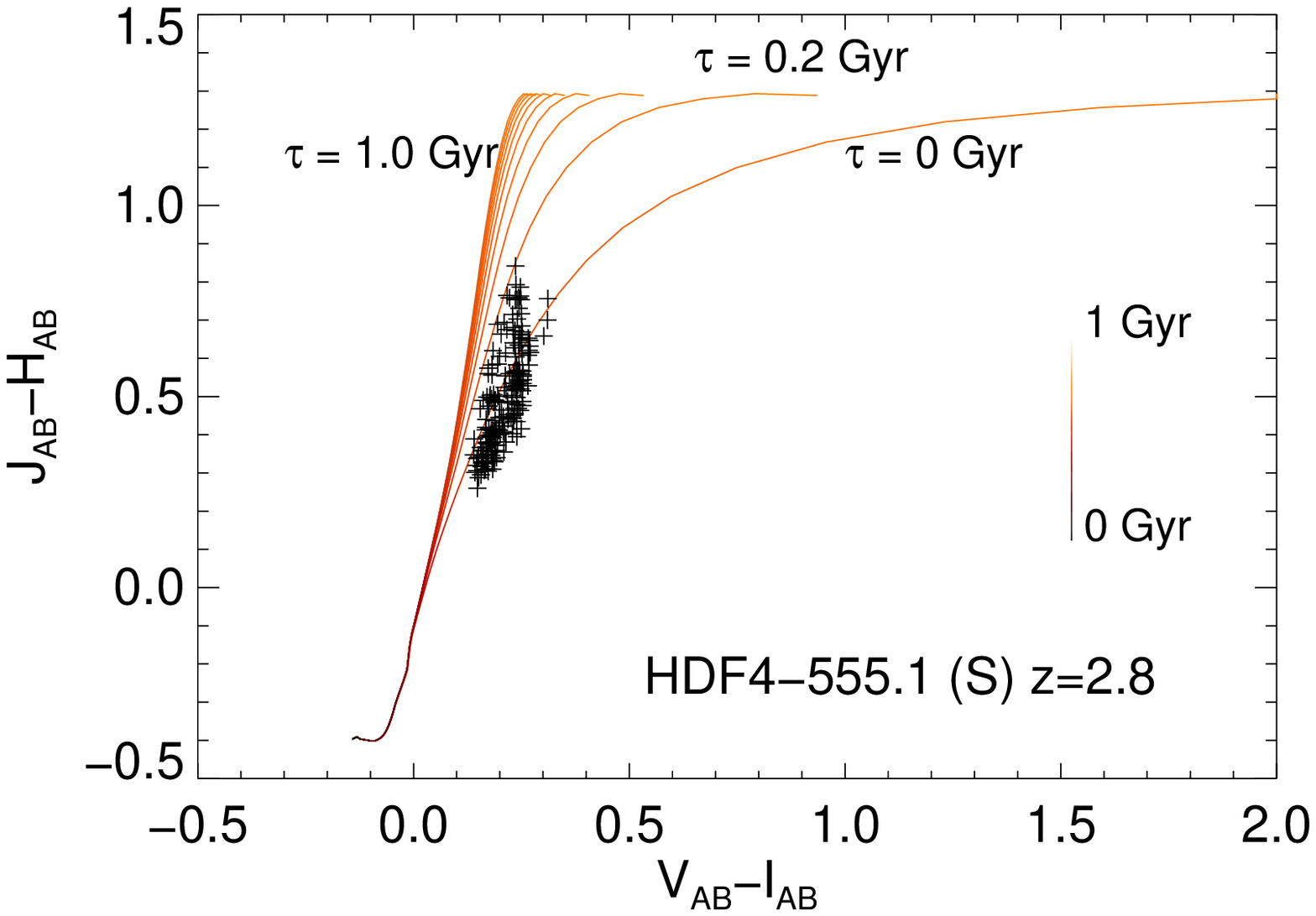}
\caption{Spatially-resolved colors of the
northern and southern components of the chain galaxy called ``the Hot
Dog'' (HDF\,4-555.1; Fig.~\ref{fig:im430and555}). The northern and
southern components exhibit subtly 
different colors, attributable to either different stellar populations
or non-uniform dust extinction.  Adopting the approach of Abraham
(1997), we also plot the evolution in the $(V-I)$ and
$(J-H)$ colors with time for a Salpeter IMF and an
exponentially-decaying star formation rate, with $e$-folding times
ranging from 0.1\,Gyr to 1\,Gyr. At $z=2.8$, $(J-H)$ straddles the
age-sensitive 4000\,\AA\ break.}
\label{fig:spat_res_hotdog}
\end{figure}

Once we correct for different resolutions of NIC\,3 and WFPC\,2 (through
``PSF matching''), we can use
the spatially-resolved colors to study different stellar
populations and/or dust-reddening within a galaxy (see
Figs.~\ref{fig:spiralSEDs}\,\&\,\ref{fig:spat_res_hotdog}).
Some of the galaxies which have the same appearance at all wavelengths
fall outside the traditional Hubble tuning-fork diagram, but instead
belong to new morphological groups, such as chain galaxies
(Fig.~\ref{fig:im430and555}; Cowie, Hu \& Songaila 1995), tadpoles (van
den Bergh et al.\ 1996) and bow-shock systems
(Fig.~\ref{fig:im430and555}).

\subsubsection{Barred Spirals:}

Our data can also address the evolution of galactic bars: it has been
claimed that at faint magnitudes, the fraction of barred spirals in the
optical HDFs declines rapidly (van den Bergh et al.\ 1996, Abraham et
al.\ 1999). If this is a truly evolutionary effect, then it has great
significance for the physics of disk formation: bars are supported by
disk self-gravity, so the implication would be either that at high-$z$
the halo mass dominates that of the disk, or there are significant
random motions in the stellar orbits (Ostriker \& Peebles 1973).

\begin{figure}[ht]
\plottwo{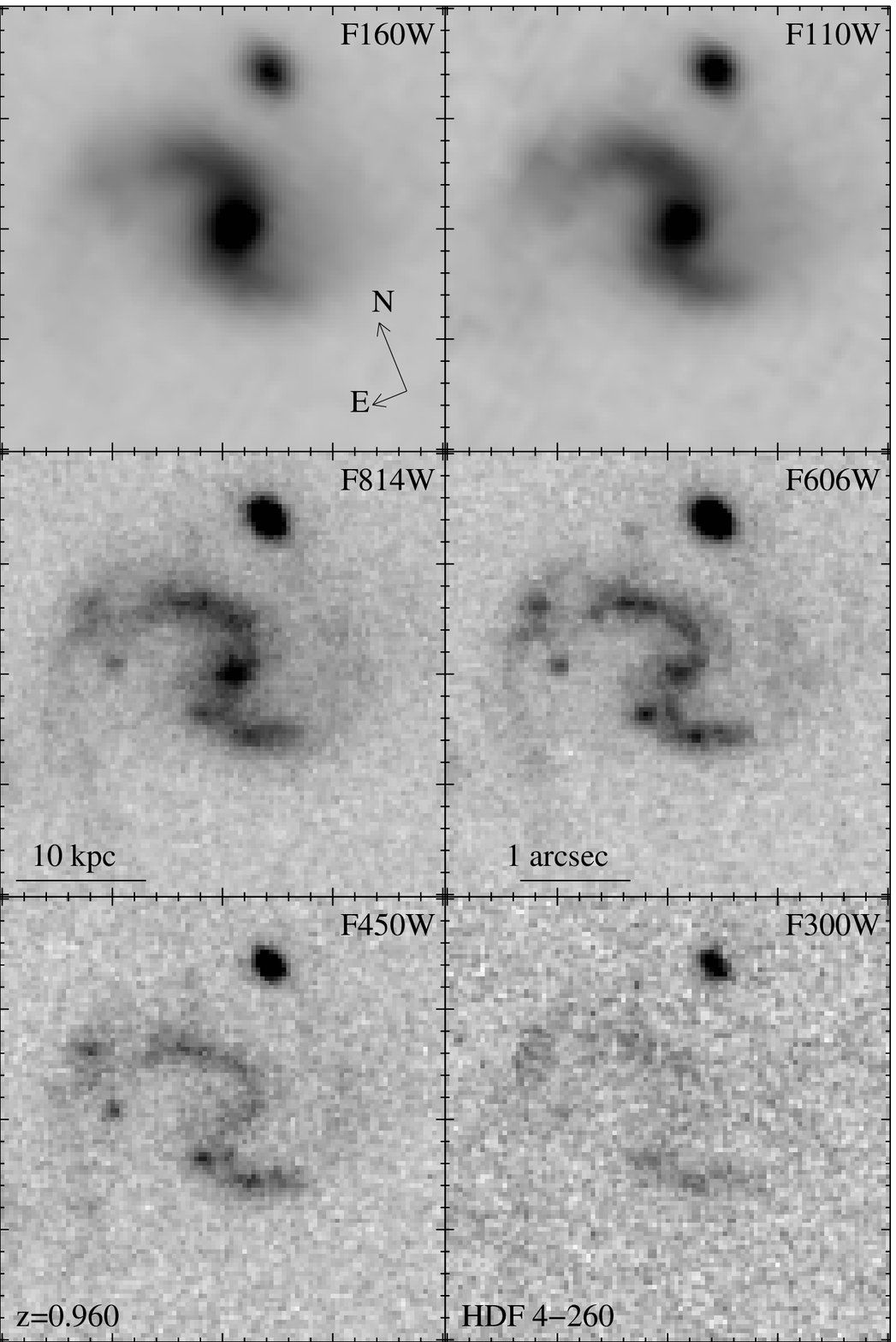}{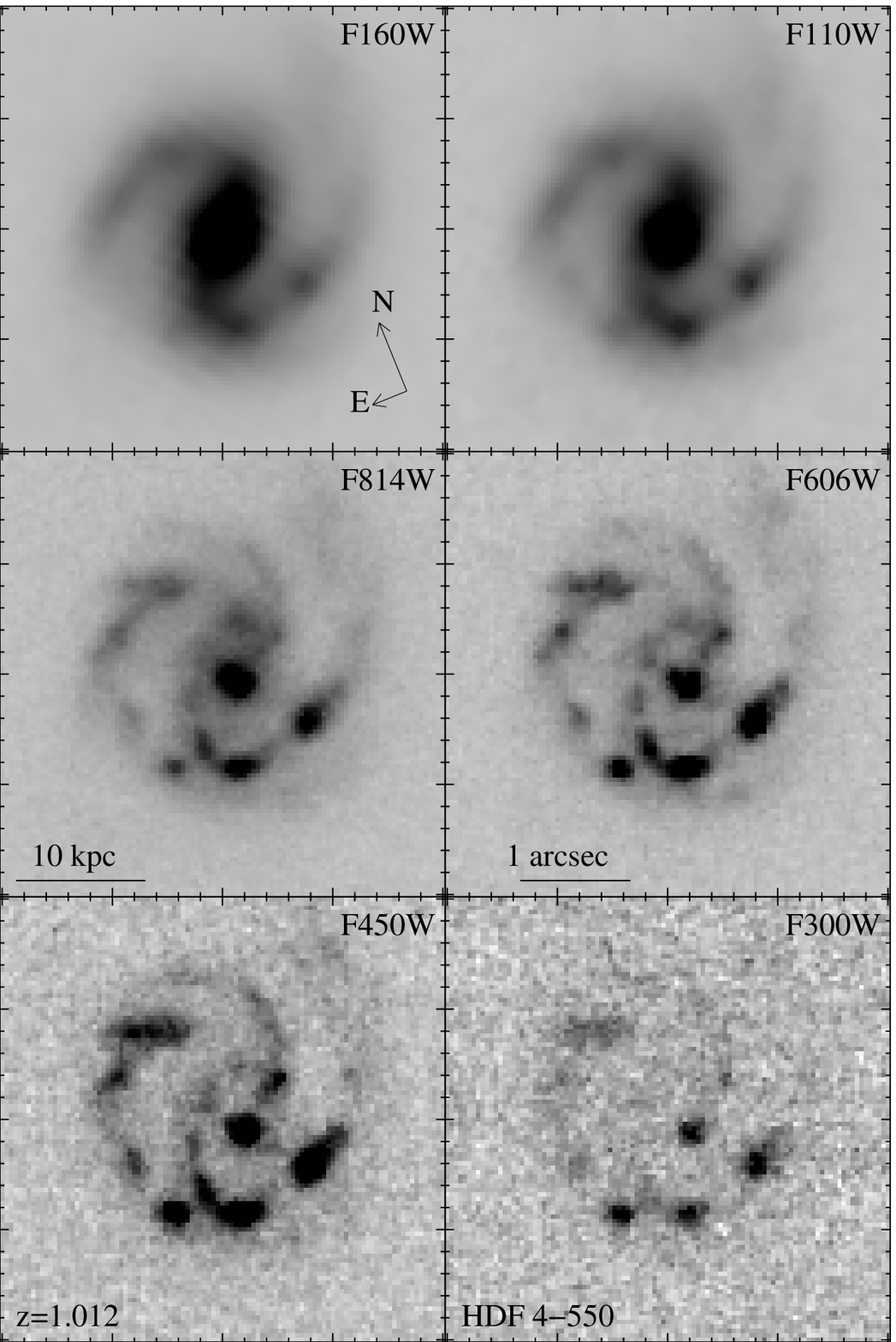}
\caption{The left panel
shows the only optically-selected barred spiral in HDF-North (van den
Bergh et al.\ 1996), and
this seems to be through chance alignment of a swath of young stars with
the approximate axis of the true bar. The
galactic bar in the spiral displayed in the right panel is only
recognizable at infrared-wavelengths -- at its redshift of $z\approx
1$, the optical wavebands only sample the rest-UV, where the older \&
redder bulge/bar stellar populations are not prominent, with the light
of young stars (e.g., the H{\scriptsize II} regions) dominating. }
\label{fig:barred}
\end{figure}

However, when the spirals are imaged in the near-IR, many are revealed
to have bars which are absent in the WFPC\,2 bands
(Fig.~\ref{fig:barred}): the bars have similar colors to the bulges
(dominated by older, cooler, redder stars).  It appears that
morphological $k$-correction effects for the higher-$z$ spirals cause
the apparent decline in optically-selected barred spirals at fainter
magnitudes.  From the small-number statistics of spirals in the
GTO-NICMOS field, there is no significant evolution in the incidence of
galactic bars.

\section{Conclusions}

Some Hubble tuning-fork galaxies only reveal
their true morphology in 
near-IR. This is particularly so for galaxies with a large dispersion in
stellar ages and spatially-distinct stellar populations, such as
in spiral galaxies.
However, such galaxies which undergo a
morphological metamorphosis
from the WFPC\,2 to NIC\,3 images are rare; most retain the same
appearance in all wavebands, or are too compact for the structural
parameters to be determined.
Once the morphological $k$-corrections have been
accounted for, it 
appears that the fraction of true irregulars does increase at
faint magnitudes/high-$z$.
Finally, the deep near-IR data shows that there is no significant
evolution in the incidence of barred spirals with redshift: their
apparent scarcity in the optical is a band-shifting effect on the older
stellar population of their bars.
A more detailed description of this work is given in Bunker, Spinrad \&
Thompson (1999).

\acknowledgments

I wish to thank my collaborators on this program, Hyron Spinrad and
Rodger Thompson. We are grateful to Ray Weymann and Lisa
Storrie-Lombardi at OCIW for organizing an enjoyable and timely workshop
on photometric redshifts, and thank Daniel Stern, Leonidas Moustakas and
Mark Dickinson for useful discussions.  A.J.B.\ acknowledges by a NICMOS
postdoctoral research fellowship, supported in part by NASA grant
NAG\,5-3043. The observations were obtained with the NASA/ESA Hubble
Space Telescope operated by the Space Telescope Science Institute managed
by the Association of Universities for Research in Astronomy Inc.\ under
NASA contract NAS\,5-26555.

\end{document}